\documentstyle[aps,epsf,epsfig,rotate]{revtex}
\newcommand \ee{\end{equation}}
\newcommand \be{\begin{equation}}
\def\bi{\bibitem}
\newcommand \bea {\begin{eqnarray} \nonumber }
\newcommand \eea {\end{eqnarray}}
 \def\(({\left(}
 \def\)){\right)}
\def\[[{\left[}
\def\]]{\right]}
\def\bi{\bibitem}

\def\a{\alpha}

\begin{document}

\title{Dynamics within metastable states in a mean-field spin glass}
\author{A. Barrat, R. Burioni and M. M\'ezard}

\address{Laboratoire de Physique Th\'eorique de l'Ecole
Normale Sup\'erieure\footnote{Unit\'e propre du CNRS, associ\'ee \`a
 l'Ecole Normale Sup\'erieure et \`a l'Universit\'e de
Paris Sud}, 24 rue Lhomond, 75231 Paris Cedex 05, France}
\maketitle
\begin{abstract}
In this letter we present a dynamical study of the structure of
metastable states (corresponding to TAP solutions) in a
mean-field spin-glass model. After reviewing known results
of the statical approach, we use dynamics: starting
from an initial condition thermalized at a temperature between
the statical and the dynamical transition temperatures,
we are able to study the relaxational dynamics
within metastable states and we show that they are
characterized by a true breaking of ergodicity and exponential relaxation.
\end{abstract}

\begin{center}
LPTENS preprint 95/50
\end{center}


Submitted to: Europhysics Letters
\vskip 1cm

The recent developments in the theory of spin glass dynamics
\cite{mm_statphys} have made clearer the similarity of
behaviour in
spin glasses and in glasses \cite{fraher,bocukume}. In
this context it seems at the moment that a certain
category of spin glasses, those which are described
by a so called one step replica symmetry breaking
(RSB) transition \cite{mpv}, are good candidate models for a
mean field description of the glass phase \cite{kithu,giorgio_verre}.
In these systems the presence of metastable states
generates a purely dynamical transition
(which is supposed to be rounded in finite dimensional
systems \cite{kithu,giorgio_verre}) at a temperature $T_d$ higher than the
one obtained within a theory of static equilibrium, $T_s$.

The spherical p-spin spin glass introduced in
\cite{crisom,crihorsom} is an interesting example of this category.
It is a simple enough system in which the metastable
states can be defined and studied by the TAP method \cite{tap}.
In this paper we want to provide a better understanding
of these metastable states, using a dynamical point of view.
We shall show the existence of
a true ergodicity breaking such that these metastable
states, in spite of being excited states with
a  finite excitation free energy per spin,
are actually dynamically stable even at temperatures
above $T_d$.

The spherical p-spin spin glass describes $N$ real spins
$s_i, \ i\in\{1,...,N\}$ which interact through the Hamiltonian:
\begin{equation}
  H(\sigma) =
     - \sum_{1\leq i_1<\cdots<i_p\leq N}\, J_{i_1,\ldots,i_p}\,
       s_{i_1}\cdots s_{i_p}
\label{ham}
\end{equation}
together with the spherical constraint on the spins:
$
\sum_{i=1}^N s_i^2 = N.
$
The couplings are gaussian, with zero mean and variance
$p!/(2N^{p-1})$.
In the $p>2$ case it shows an interesting
spin glass behaviour, simple enough to allow for detailed
analytical treatment.

In the static approach, one
describes the properties of the Boltzmann probability
distribution of this system. The replica method  shows
the existence of a static transition with a one step RSB
at temperature $T_s$ \cite{crisom}. This transition
reflects the fact that, below $T_s$, the Boltzmann
measure is dominated by a few number of pure states,
a scenario which is well known from the random
energy model \cite{rem}.

 Staying within a static framework,
the TAP approach \cite{kuparvir,crisom94} provides
some more insight into the physical nature of this system.
The TAP equations can be derived through a variational principle
on the local magnetizations $m_i= <s_i>$, from a free energy
$f\((\{m_i \} \))$ which is best written in terms of radial
and angular variables, $q$ and  $\hat{s}_i$ (with
$m_i = \sqrt{q} \hat{s}_i$), in the form \cite{kuparvir}:
\be
f\((\{m_i \} \))=q^{p/2} E^0\((\{ \hat{s}_i \} \))
 - \frac{T}{2}\ln(1-q) -
\frac{1}{4T}[(p-1)q^p - pq^{p-1} +1] \ ;
\label{ftap}
\end{equation}
where the angular energy is:
\be
E^0 \((\{ \hat{s}_i \} \))
\equiv - \sum_{1\leq i_1<\cdots<i_p\leq N}\, J_{i_1,\ldots,i_p}\,
       \hat{s}_{i_1}\cdots \hat{s}_{i_p} \ .
\ee
At zero temperature the TAP states
are just unit vectors which minimize the
angular energy $E^0$. There actually exist such states for
 $E^0 \in [E_{min}, E_c=-\sqrt{2(p-1)/p}$].
Denoting by $\hat{s}_i^\alpha$ one zero temperature
state, of
energy $E^0_\alpha$, it gives rise at finite
temperature $T$ to one TAP state $\alpha$ given by:
\be
m_i^\alpha= \sqrt{q \((E^0_\alpha,T \))} \hat{s}_i^\alpha \ ,
\ee
where $q(E,T)$ is the largest solution of the equation:
\be
(1-q) q^{p/2-1} = T \(( {-E-\sqrt{E^2-E_c^2} \over p-1} \)) \ .
\label{qtap}
\ee
The free energy of this state, $f_\a$, at temperature
$T$, is obtained by inserting
in the TAP free energy (\ref{ftap}) the corresponding
values of the angular energy, $E^0=E^0_\a$ and of the self
overlap, $q = q_\a \equiv q \((E^0_\alpha,T \))$.
The corresponding energy is:
\be
E_\alpha=q_\alpha^{p/2} E^0_\alpha - \frac{1}{2T}
[(p-1)q_\alpha^p - pq_\alpha^{p-1} +1] \ .
\label{etap}
\ee
When changing the temperature, one can follow the
metastable states which keep the same angular direction; their
order in free energy or energy, at fixed T, is the same as their
order in $E^0$.
When raising $T$,
a state disappears at a temperature $T_{max}(E^0)$ (where
equation (\ref{qtap}) ceases to have solutions). $T_{max}(E^0)$ is
a decreasing function of $E^0$;
the most excited states, with $E^0=E_c$,
disappear first at $T_{max}(E_c)$,
and the lowest
at $T_{max}(E_{min}) \equiv T_{TAP}$. Above $T_{TAP}$,
the only remaining state is the paramagnetic one with $q=0$ and
free energy $F_{para}=-1/(4 T)$.

To complete the description of metastable states at any temperature,
one only needs the density
of states $\rho(E^0)$ with an angular energy $E^0$. This
has been computed in \cite{crisom94}; the multiplicity is exponentially
large, giving a finite complexity density $s_c^0(E^0)$, defined
as:
\be
s_c^0(E^0)= \lim_{N \to \infty} {\log \rho(E^0) \over N} \ .
\ee
The complexity at finite temperature is easily deduced from
this $s_c^0$. We shall denote by $S_c(f,T)$ the logarithm of
the number of TAP states at free energy $f$ and temperature $T$.
The Boltzmann partition function can  then be approximated as the
sum over all TAP solutions:
\be
Z= \int df \exp \(( -{ (f-T S_c(f,T) ) \over T } \))
\ee
which can be evaluated at large $N$ by a saddle point
method. At temperatures $T>T_d$, with
$T_d=\sqrt{p(p-2)^{p-2}(p-1)^{1-p}/2}$, the Boltzmann measure
is dominated by the paramagnetic state $q=0$.
At any $T \in [T_s, T_d]$,  the Boltzmann
measure is dominated by a class of TAP solutions, those
of free energy $f=f_{eq}(T)$. Because of their extensive complexity,
this gives for the  total equilibrium free energy:
\be
f_{tot} \equiv -T \ln (Z)= f_{eq}(T)-T S_c\((f_{eq}(T),T\)) \ .
\ee
The computation of $f_{eq}$ is easily done \cite{crisom,monasson2}.
One finds that $f_{tot}$ is {\it equal to} the paramagnetic free
energy in this range. Below $T_s$ the lowest lying TAP states
dominate the Boltzmann measure, leading to RSB. The
situation is summarized in figure 1.
Compared to usual phase transition, the situation is
complicated by the
existence of a finite complexity. Actually we see
 that between the two  transition temperatures $T_s$ and $T_d$, the
situation is unclear: the total equilibrium free energy seems
to get two equal contributions, from the paramagnetic state
and from a bunch of TAP solutions
with non-zero $q$. One can wonder if there is a phase coexistence,
or simply a problem of double counting in the TAP approach.
This issue, which is an important one
if one aims at understanding the finite dimensional behaviour
of this type
of systems \cite{giorgio_verre}, can in fact be clarified within
a dynamical approach as we now show. Let us also mention that some
purely static approaches also carry relevant
information on related issues \cite{monasson,franzpar95}.

\begin{center}
\parbox{6cm}{
\epsfig{figure=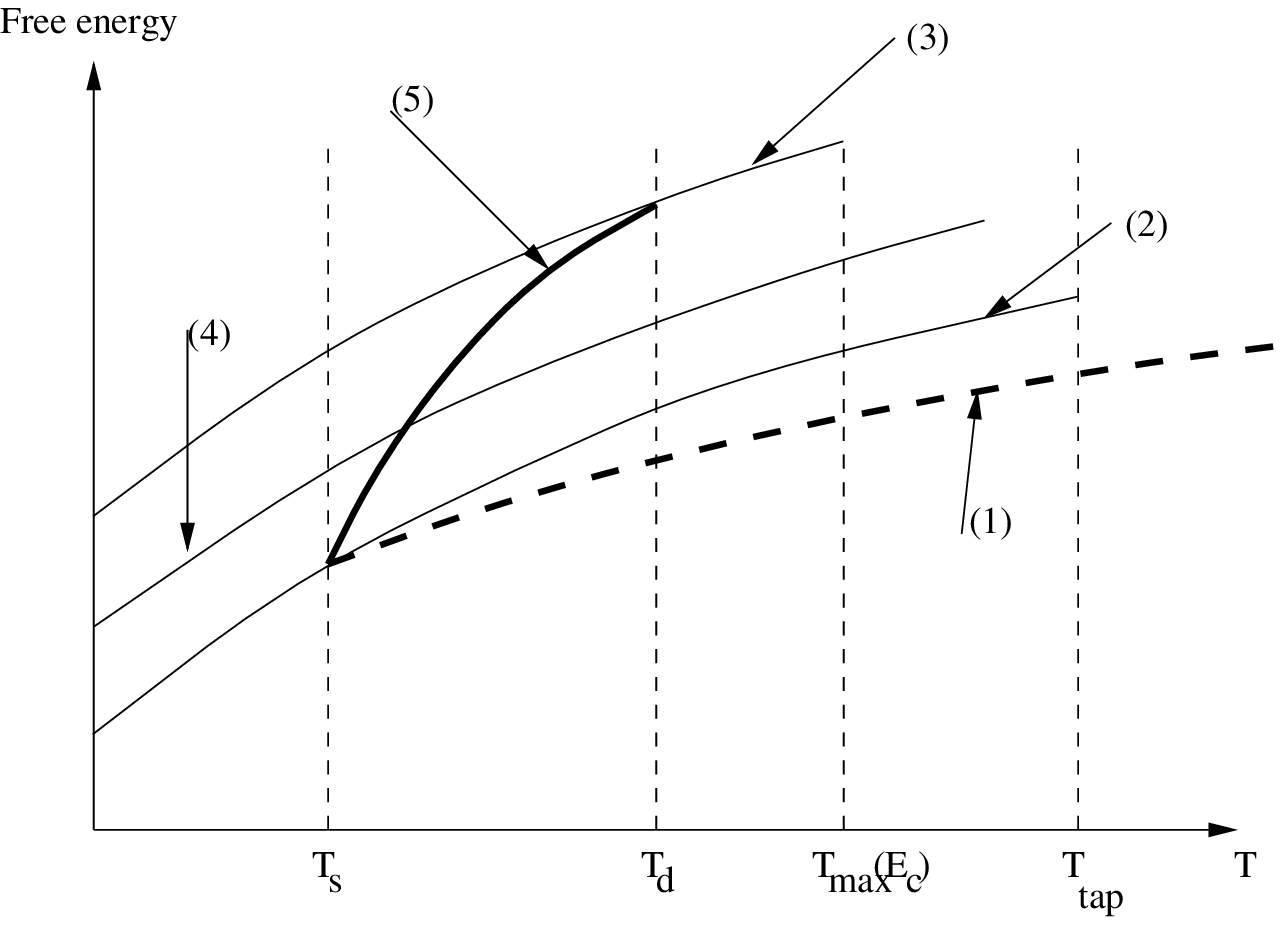,width=6cm,angle=-90}}
\\[.3cm]
{\footnotesize Free energy versus temperature; (1): free energy of the
paramagnetic solution for $T > T_d$, $f_{tot}$  for $T <  T_d$;
(2): free energy of the lowest TAP states, with zero temperature
energy $E_{min}$; (3): free energy of the highest TAP states, corresponding
to $E_c$; (4): an intermediate value of $E_0$ leads to an
intermediate value of $f$ at any temperature; (5): $f_{eq}(T)$; the
difference between curves (5) and (1) gives the complexity
$T S_c\((f_{eq}(T),T\))$.}
\end{center}

The TAP structure of states has never been explored dynamically:
indeed, the usually studied out of equilibrium dynamics of the
spherical p-spin model starts from a random configuration,
and never goes below the threshold corresponding to the upper
TAP solutions.
This process has been studied in \cite{cukuprl}: an interesting
aging behaviour has been found
at temperatures $T<T_d$, but the energy density of the system
only goes asymptotically to the one of the highest
TAP states (the threshold states with angular energy $E^0=E_c$).
Hence, it is impossible to explore TAP states via this kind of dynamics.

Here we will use a different approach for
the dynamics \cite{franzpar95}, where we start from a spin configuration which
is picked up from a Boltzmann distribution at temperature
$T'$, and then let the system relax at temperature T.
We shall concentrate on the case where $T' \in [T_s,T_d]$,
which means that our initial configuration will belong
to the TAP states with free energy $f_{eq}(T')$. This will
lead to the study of the
relaxation {\it inside} one TAP state.

The relaxational dynamics at temperature $T$
is given by the Langevin equation:
\begin{equation}
\frac{ds_i(t)}{dt} = -\frac{\partial H}{\partial s_i}
- \mu(t)s_i(t) + \eta_i(t)
\end{equation}
where $H$ is the Hamiltonian (\ref{ham}), $\mu$ is the Lagrange
multiplier implementing the spherical constraint, and $\eta_i$
is a gaussian white noise with zero mean and variance $2T$.
The dynamics
is described by the behaviour of two-times
correlation and reponse functions defined by:
\begin{equation}
C(t,t')=\frac{1}{N} \sum_{i=1}^N \overline{<s_i(t)s_i(t')>} , \ \
r(t,t')=\frac{1}{N} \sum_{i=1}^N
\frac{\partial \overline{<s_i(t)>}}{\partial h_i(t')},
\end{equation}
where $<.>$ is a mean over the thermal noise, and an overline denotes
a mean over the coupling constants.

Using the usual field theoretical techniques for out
of equilibrium dynamics \cite{sompzip}, in the large $N$ limit,
it is possible to study the dynamics at temperature $T$,
starting from a Boltzmann measure at temperature $T'$. In order to implement
this initial -sample dependent- measure, it is
necessary to introduce replicas \cite{houghton83,franzpar95} and
to write dynamical equations for two-times overlaps between replicas
$C^{ab}(t,t')=\overline{<s^a(t)s^b(t')>}$,
$a$ and $b$ being replica indices.
The obtained equations differ from the usual out of
equilibrium ones (corresponding to $T'=\infty$ \cite{cukuprl})
 by terms involving a coupling to
the initial configuration, i.e. $C^{ab}(t,0)$. Besides, as
noted in \cite{franzpar95}, the time evolution respects
the initial replica symmetric or RSB structure of the $C^{ab}$, i.e.
the static replica structure describing equilibrium at $T'$.

For the p-spin model with $T'>T_s$
 the initial condition is replica symmetric,
with $C^{ab}(0,0)=\delta_{ab}$. Therefore,
at all times we can write $C^{ab}(t,t')=C(t,t')\delta_{ab}$.
The obtained equations for the correlation and response functions
read\cite{franzpar95}, for any $T'> T_s$, and $t>t'$:
\bea
\mu(t) &=& \int_0^t ds\ \left[ \frac{p^2}{2} C^{p-1}(t,s)
-\frac{p(p-1)}{2} C^{p-2}(t,s) \right] r(t,s) + T \nonumber \\
&-& \frac{p}{2T'} C^{p-1}(t,0)\ (1-C(t,0))  \nonumber \\
{\partial r(t,t') \over \partial t}&=& -\mu(t)  r(t,t')
-\frac{p(p-1)}{2} \int_0^t ds  \ C^{p-2}(t,s)r(t,s)(r(t,t')-r(s,t'))
\nonumber \\
&-& \frac{p}{2T'} C^{p-1}(t,0)\ r(t,t') \nonumber \\
{\partial C(t,t') \over \partial t} &=&-\mu(t)C(t,t')
+ \frac{p}{2} \int_0^{t'} ds \  C^{p-1}(t,s)r(t',s) \nonumber \\
&-& \frac{p(p-1)}{2} \int_0^t ds  \ C^{p-2}(t,s)r(t,s)(C(t,t')-C(s,t'))
\nonumber \\
&-&\frac{p}{2T'} C^{p-1}(t,0)\ C(t,t') +
\frac{p}{2T'} C^{p-1}(t,0)\ C(t',0)
\label{eqTT}
\eea

Let us examine the situation first for $T=T'$: since we start at
equilibrium, we expect equilibrium dynamics satisfying
both time translation invariance
(TTI) and the fluctuation dissipation theorem
(FDT): $C(t,t')=C_{eq}(t-t'),\ r(t,t')=r_{eq}(t-t')$
with $r_{eq}(\tau)=- \frac{1}{T} \frac{\partial C_{eq}}{\partial \tau}$.
The equations (\ref{eqTT}),
with this Ansatz, reduce to a single equation for the evolution
of $C_{eq}(\tau)$:
\be
{\partial C_{eq}(\tau) \over \partial \tau} = -\mu_{\infty}C_{eq}(\tau)
- \frac{p}{2T} \int_0^{\tau}du\ C_{eq}^{p-1}(\tau-u)
\ \frac{\partial C_{eq}(\tau)}{\partial \tau}
\label{eqas}
\ee
where $\mu_{\infty}=T$, and $C_{eq}(0)=1$. Above $T_d$, this equation
describes the relaxation within the paramagnetic
state, with $\lim_{\tau \to \infty}C_{eq}(\tau)=0$.
Below $T_d$, the condition of dynamical stability
${\partial C_{eq}(\tau) \over \partial \tau} \le 0$
leads to a non zero
limit $C_{\infty}$ for $C_{eq}(\tau)$
\cite{crihorsom}; this limit is given by the
largest solution of:
\be
\frac{p}{2T^2}C_{\infty}^{p-2}(1-C_{\infty}) = 1
\ee
(the other non zero solution is unstable with respect to the dynamics
(\ref{eqas})).
This value is precisely the self overlap $q$ of the TAP states
reflecting the statics at $T$, i.e. with free energy $f_{eq}(T)$.
This means that, for temperatures between the statical and
the dynamical transition temperatures,
the thermalized system is trapped inside a TAP state,
and not in a paramagnetic state, for which
$C_\infty$ would be zero (as for $T>T_d$). We can also
exclude the possibility of a coexistence,
which would lead to some intermediate value: the paramagnetic state
has disappeared at $T_d$, and the Gibbs state is formed by the
bunch of TAP solutions having the suitable free
energy $f_{eq}(T)$, and a finite complexity density.

To get further insight, always starting
from a thermalized configuration at temperature $T' \in [T_s,T_d]$,
we now study the dynamics
at a temperature $T$ different from $T'$.
In our study of the
dynamical equations (\ref{eqTT}), we have found
numerically (using the type of algorithm developped
in \cite{franzmezard}) that the system
reaches after a short transient a stationnary regime
where TTI and FDT hold (see figure 2). We have then shown
analytically the consistency of such a solution.
The possibility of such a situation has already been conjectured
in \cite{franzpar95}, together with an interesting connection
to the static approaches developped in \cite{monasson,franzpar95}.

\begin{center}
\parbox{5.6cm}{
\epsfig{figure=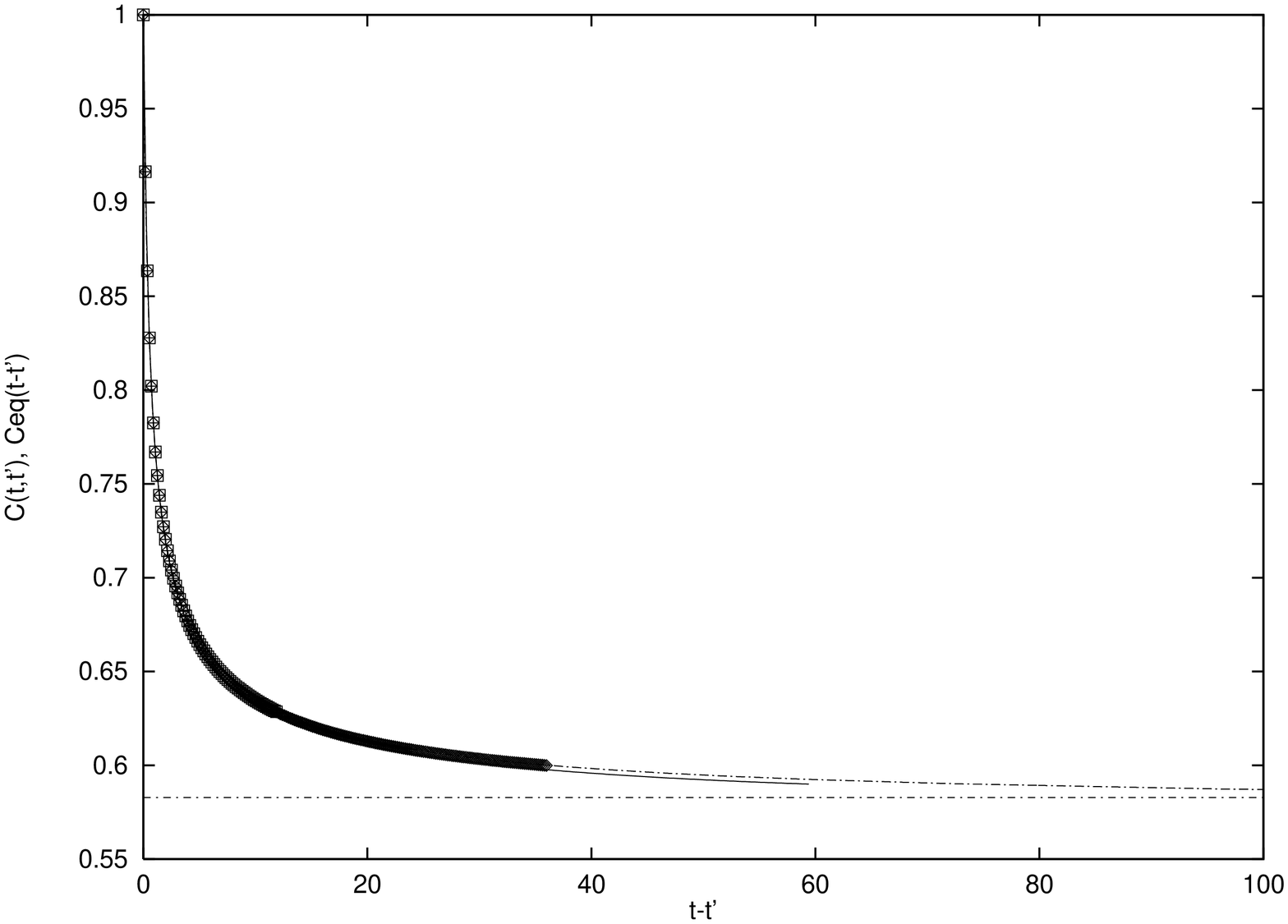,width=6cm,angle=-90}}
\\[.3cm]
{\footnotesize
p=3 model, with $T_s \approx .586$, $T_d \approx .612$;
numerical integration of equations (\ref{eqTT}) for
$T'=.605$, $T=.6$; we plot $C(t,0)$ versus t (solid line), and
$C(t,t')$ versus $t-t'$ for $t'=6,12,18,24$ (symbols);
the dotted curve is the numerical integration of (\ref{as2}),
and the dotted line is the value of  $C_\infty$ obtained by (\ref{as3}).
}
\end{center}

In order to study this solution analytically,
we introduce as previously $C_{eq}(\tau)$, $r_{eq}(\tau)$,
$C_\infty=\lim_{\tau \to \infty} C_{eq}(\tau)$,
and $l=\lim_{t \to \infty} C(t,0)$,
and obtain the equation:
\bea
{\partial C_{eq}(\tau) \over \partial \tau}&=& - \left(\mu_{\infty} +
\frac{p}{2T} C_{\infty}^{p-1} -
\frac{p}{2T'} l^{p-1}\right) C_{eq}(\tau) \nonumber \\
&+&\frac{p}{2} \int_0^{\tau}du\ C_{eq}^{p-1}(u)\ r_{eq}(\tau-u)
-\frac{p}{2T} C_{\infty}^p + \frac{p}{2T'} l^p.
\label{as2}
\eea
Besides, $\mu_{\infty}$, $C_{\infty}$ and $l$ satisfy the following
set of equations, obtained by taking $t'=0$, $t \to \infty$ in
(\ref{eqTT}), and $\tau \to \infty$ in (\ref{as2}):
\bea
\mu_{\infty}&=&T +  \frac{p}{2T} C_{\infty}^{p-1} (1-C_{\infty})
-\frac{p}{2T'} l^{p-1} (1-l) \nonumber \\
l^{p-2} &=& \frac{2T T'}{p (1-C_{\infty})} \nonumber \\
T C_{\infty} &=& \frac{p}{2T'} l^p (1-C_{\infty}) + \frac{p}{2T}
C_{\infty}^{p-1} (1-C_{\infty})^2
\label{as3}
\eea
and the energy reached dynamically at large times is:
$E_\infty = \frac{1}{2T}( C_{\infty}^p - 1 ) - \frac{l^p}{2T'}$.

It is then straightforward to check that the overlap
 $C_{\infty}$ and  the energy $E_{\infty}$ are identical to the values
characteristic of certain TAP states at the temperature $T$. These
states are precisely those obtained by following the equilibrium
TAP states at temperature $T'$ (which pick up a certain value $E^0_{T'}$
of the angular energy) to temperature $T$, by
keeping the same direction in $\hat{s}$ space, but changing the overlap
from $q(E^0_{T'},T')$ to $q(E^0_{T'},T)$.

{}From (\ref{as2}), it is possible to show that
the relaxation of $C_{eq}(\tau)$ is of the form
$\tau^{-3/2} exp(-\tau/\tau_0)$. The relaxation time $\tau_0$
can also be computed, and has a quite complicated expression
that we do not reproduce here.
It diverges for the highest TAP states (corresponding to $E^0=E_c$).
Of course, this exponential relaxation can only happen
as long as the followed TAP solution
still exists at temperature $T$: if $T$ becomes larger than
$T_{max}(E^0_{T'})$, we observe a fast relaxation
to the paramagnetic state, with $C_\infty=l=0$.

We have thus shown that the TAP solutions are real states,
corresponding to a full breaking of ergodicity: starting
within a TAP state (which can be achieved by our trick of
using
thermalized initial conditions at a temperature $T'$), one
relaxes within this state with a finite relaxation rate, and one
can even follow this state when changing the temperature.
 Besides, the Gibbs measure below the
dynamical transition is made of a superposition of TAP states,
which are different ergodic components, totally separated
from each other in the dynamical evolution.
The paramagnetic solution, valid above $T_d$, disappears
at $T_d$. Note that the way in which this
occurs is not clear, and we leave this open question,
which is crucial for a better understanding of aging dynamics,
for future work.
Some TAP states exist as independent ergodic components even
at temperatures $T \in [T_d,T_{TAP}]$. They are not seen in the
usual dynamics because they are difficult to find:
starting from random initial conditions one stays in
the big paramagnetic ergodic component. If one succeeds
in starting within a TAP state, one stays
within this state even by rising the temperature above $T_d$ (but
below the $T_{max}$ of this state).
One should notice that the
usual dynamics at a temperature below $T_d$, starting
from a random configuration, only leads to
a ``weak ergodicity breaking'' \cite{bouchaud,cukuprl},
where the self overlap vanishes at very large time
differences (much larger than the waiting time).
This is explained\cite{cukuprl,kurlal}
by the fact that the system, which was initially
in the (infinite temperature) paramagnetic state, does
not find any TAP state in a finite time, but stays
at energy density $O(1)$ (going to zero as $t$ goes to infinity)
above the threshold.
On the contrary, there is no sign
of aging when one starts within a TAP state. This is in agreement with
some recent intuitive scenarios for aging \cite{kurlal,barmez}.

It is a pleasure to thank J. Kurchan and R.Monasson for some very useful
discussions and suggestions.


\begin{references}
\bi{mm_statphys} For a short introduction and references, see
M. M\'ezard, lecture at Statphys19, LPTENS 95/35;
\bi{fraher} S. Franz and J. Hertz, Phys. Rev. Lett. {\bf 74} (1995) 2114;
\bi{bocukume} J.P Bouchaud, L. Cugliandolo, J. Kurchan and M. M\'ezard,
preprint LPTENS 95/47, condmat 9511042;
\bi{mpv}M. M\'ezard, G. Parisi and M.A. Virasoro, "Spin glass theory
and beyond",  (World Scientific, Singapore 1987);
\bi{kithu} T. R. Kirkpatrick, D. Thirumalai, Phys. Rev. {\bf B36} (1987) 5388,
T. R. Kirkpatrick, D. Thirumalai, P. G. Wolynes,
Phys. Rev. A {\bf 40} (1989) 1045 and references therein;
\bi{giorgio_verre} G. Parisi, preprint condmat-9412034;
\bi{crisom} A. Crisanti, H.-J. Sommers, Z. Physik {\bf B 87} (1992)
341;
\bi{crihorsom} A. Crisanti, H. Horner, H.-J. Sommers, Z. Physik {\bf B 92}
(1993) 257;
\bi{tap} D.J. Thouless, P.W. Anderson, R.G. Palmer, Phil. Mag. {\bf 35}
(1977) 597;

\bi{rem} B. Derrida, Phys. Rev. Lett. {\bf 45} (1980) 79;
D. Gross and M. M\'ezard, Nucl. Phys. {\bf B240} (1984) 431;

\bi{kuparvir} J. Kurchan, G. Parisi, M. A. Virasoro, J.Phys. I
France {\bf 3} (1993) 1819;

\bi{crisom94} A. Crisanti, H.-J. Sommers, J. Physique I {\bf 5} (1995) 805;
\bi{monasson} R. Monasson, Phys. Rev. Lett. {\bf 75} (1995) 2847;
\bi{monasson2} R. Monasson, in preparation;
\bi{cukuprl}  L.F. Cugliandolo, J. Kurchan, Phys.Rev.Lett. {\bf 71} (1993) 173;
\bi{sompzip}
H.~Sompolinsky and A.~Zippelius, Phys.Rev.Lett. {\bf 47} (1981) 359;
  Phys.Rev. {\bf B25} (1982) 6860;
\bi{houghton83} A. Houghton, S. Jain, A.P. Young, Phys. Rev. {\bf B 28}
(1983) 290;

\bi{franzpar95} S. Franz, G. Parisi, preprint condmat 9503167;

\bi{franzmezard} S. Franz, M. M\'ezard, Physica {\bf A 210} (1994) 48;
\bi{bouchaud}J. P. Bouchaud, J. Physique I 2  (1992) 1705;
\bi{kurlal} J. Kurchan and L. Laloux, preprint condmat 9510079;
\bi{barmez} A. Barrat, M. M\'ezard, J. Physique I {\bf 5} (1995) 941;
\end{references}
\end{document}